\documentstyle[12pt]{article}

\input BoxedEPS.tex
\SetTexturesEPSFSpecial
\HideDisplacementBoxes

\begin{document}

\title{The Allegro gravitational wave detector - data acquisition and analysis}
\author{E.~Mauceli, Z.~K.~Geng, W.~O.~Hamilton, W.~W.~Johnson, \\
 S.~Merkowitz, A.~Morse, B.~Price and N.~Solomonson}

\date{\today}
\maketitle

\begin{abstract} We discuss the data acquisition and analysis procedures used on the
Allegro gravity wave detector, including a full description of the filtering used for
bursts of gravity waves.  The uncertainties introduced into timing and signal strength
estimates due to stationary noise are measured, giving the windows for both quantities in
coincidence searches.

\end{abstract}

\section{Introduction}

For the past ten years there has been little doubt that gravitational waves exist
~\cite{taylor} .  The extended series of measurements on the orbital decay of the binary
pulsar have made it clear that angular momentum is radiating away from this system in
agreement with the original predictions of Einstein.  but, the larger goal of direct
detection of the waves and the development of such detection into gravitational wave
astronomy still remains.  Direct detection of gravitational radiation is a challenging
experimental and technological problem.  The current state of gravitational wave
experimentation will allow detectors to record any predicted event that occurs within our
galaxy and the technology is at hand to allow experimentalists to record events from
remote galaxies.  We report here on the data acquisition and analysis procedures used for
the Allegro gravity wave detector, including the design of the optimal filter for burst
signals and quantifying the uncertainties in estimating arrival times and signal
strengths.

\section{The Detector}

\begin{figure}
	\centerline{\BoxedEPSF{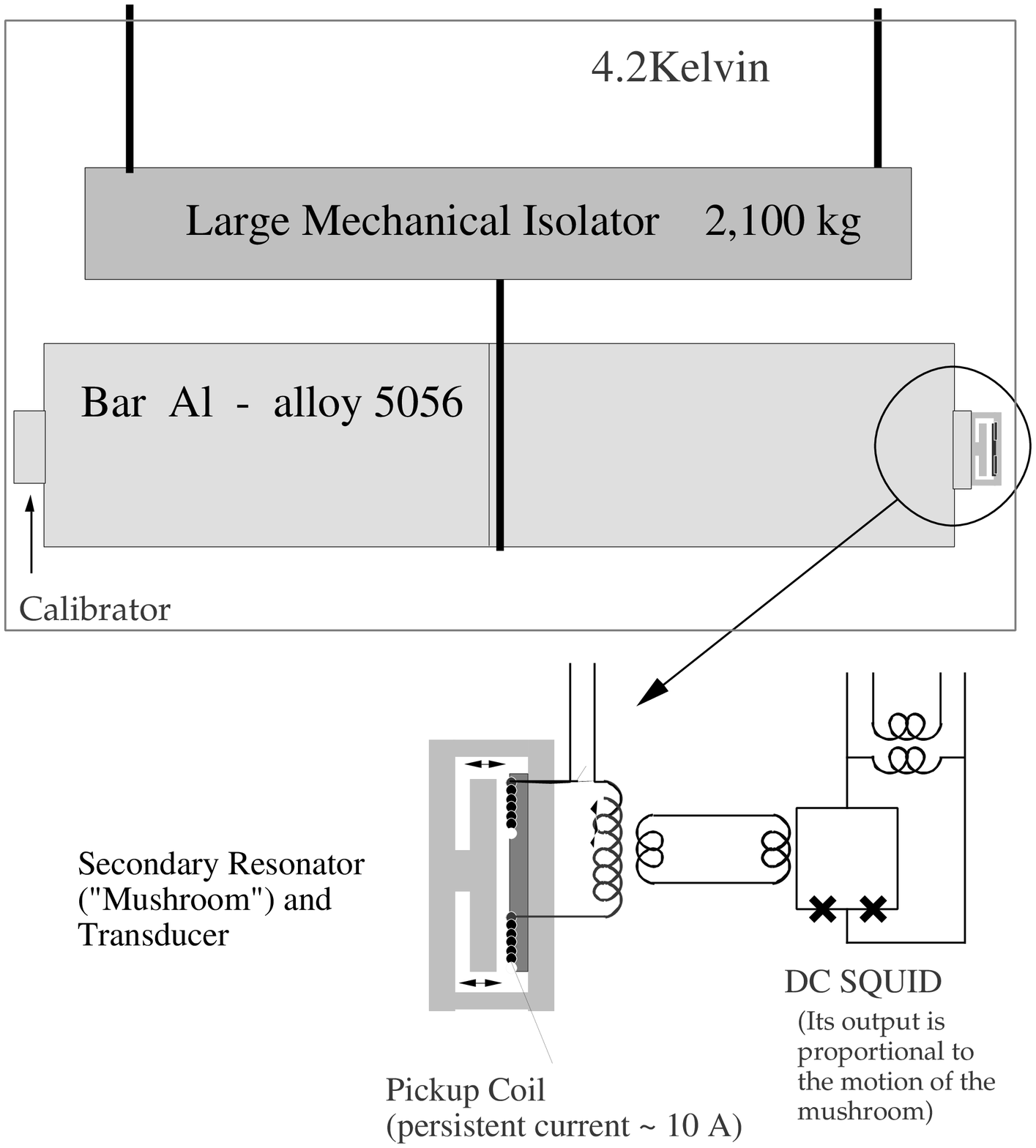 scaled 500}}
 \caption{The schematic of the Allegro antenna.} 
 \label{crossec}
\end{figure}

Allegro is located in the Physics Building at Louisiana State University in Baton Rouge,
Louisiana  ($30^{\circ} \, 25' \: N, \: 91^{\circ} \, 10' \: W
$).  It consists of a resonant bar equipped with a resonant inductive transducer and a dc
SQUID amplifier all cooled to 4.2 K.  It was operational from June 1991 until January of
1995 with a duty cycle approaching 95\% and an average noise temperature (defined in
Sec.~\ref{sec:datanaly}) less than 6 mK.  Figure~\ref{crossec} shows a schematic of the
antenna. 

\subsection{The Bar}

The Allegro detector was designed to look for pulses of gravity waves such as those from
the collapse of a massive star.  Theoretical models (although varying widely in waveform
and strength estimates) predict that stellar collapse to a neutron star or black hole
would produce a burst of gravitational radiation with a duration on the order of
milliseconds at frequencies near 1 kHz.  For a resonant mass detector, a passing gravity
wave deposits momentum into a massive elastic body, changing the amplitude and phase of
the existing vibrational normal mode motion. The elastic body in the Allegro detector is
a cylinder of aluminum alloy 5056, 60 cm in diameter and 300 cm in length. It has a
physical mass of 2296 kg.  Its first longitudinal normal mode is at 913 Hz. All cylinder
detectors are most sensitive to signals propagating in a direction perpendicular to the
bar axis.  The bar is oriented perpendicular to the plane of the great circle on the
earth that passes through Geneva, the location of the Rome Explorer antenna, and midway
between Baton Rouge, LA and Stanford, CA.  This orientation results in the axis of
Allegro being directed along a line
$40^{\circ} \, 24'$ west of North.  The Explorer detector of the University of Rome is
perpendicular to the same great circle and as a result is parallel to Allegro.  This means
that a gravity wave should deposit the same amount of momentum into each of the
detectors.  
 
\subsection{The Transducer}

Attached to one end of the bar is a smaller ``mushroom'' resonator resonant at the same
frequency as the bar, thus making a two-mode system of coupled oscillators (referred to
as the ``antenna'') ~\cite{paik_phd74,nb1_transd}.  The mass of the resonator is  small
enough so that the effects of a passing gravity wave on it are ignored.  Facing the
mushroom resonator but attached firmly to the bar is a superconducting pick-up coil with
a persistent supercurrent.  The distance between the coil and the resonator is therefore
proportional to the distance between the bar and resonator. Oscillations of the mushroom
resonator change the inductance of the pick-up coil, modulating the flux through it.  A dc
SQUID\footnote{Biomagnetic Technologies, inc. 4174 Sorrento Valley Blvd.  San Diego, CA
92121} converts the changing flux to a voltage.

\subsection{The Calibrator}

An off-resonant capacitive transducer, the calibrator, is attached to the bar at the 
opposite end from the inductive transducer.  Voltages applied to the capacitor applied
forces to the antenna, which we used for a number of tasks.  The calibrator was used to
actively dampen the mode Q's to shorten the recovery time after large excitations and to
cancel positive feedback on the antenna produced by the SQUID ~\cite{nb1_transd}.  Under
normal operating conditions the calibrator was used to excite the antenna at a frequency
of 865.00 Hz, far removed from either of the modes.  This ``continuous systems test''
provides a powerful tool for checking on the health of the detector.  The calibrator was
also used to provide burst signals to the antenna allowing the detector to be calibrated
and allowing a study of the effects of noise on signal detection to be made.  

\subsection{The Antenna Model}

\begin{figure}
  	\centerline{\BoxedEPSF{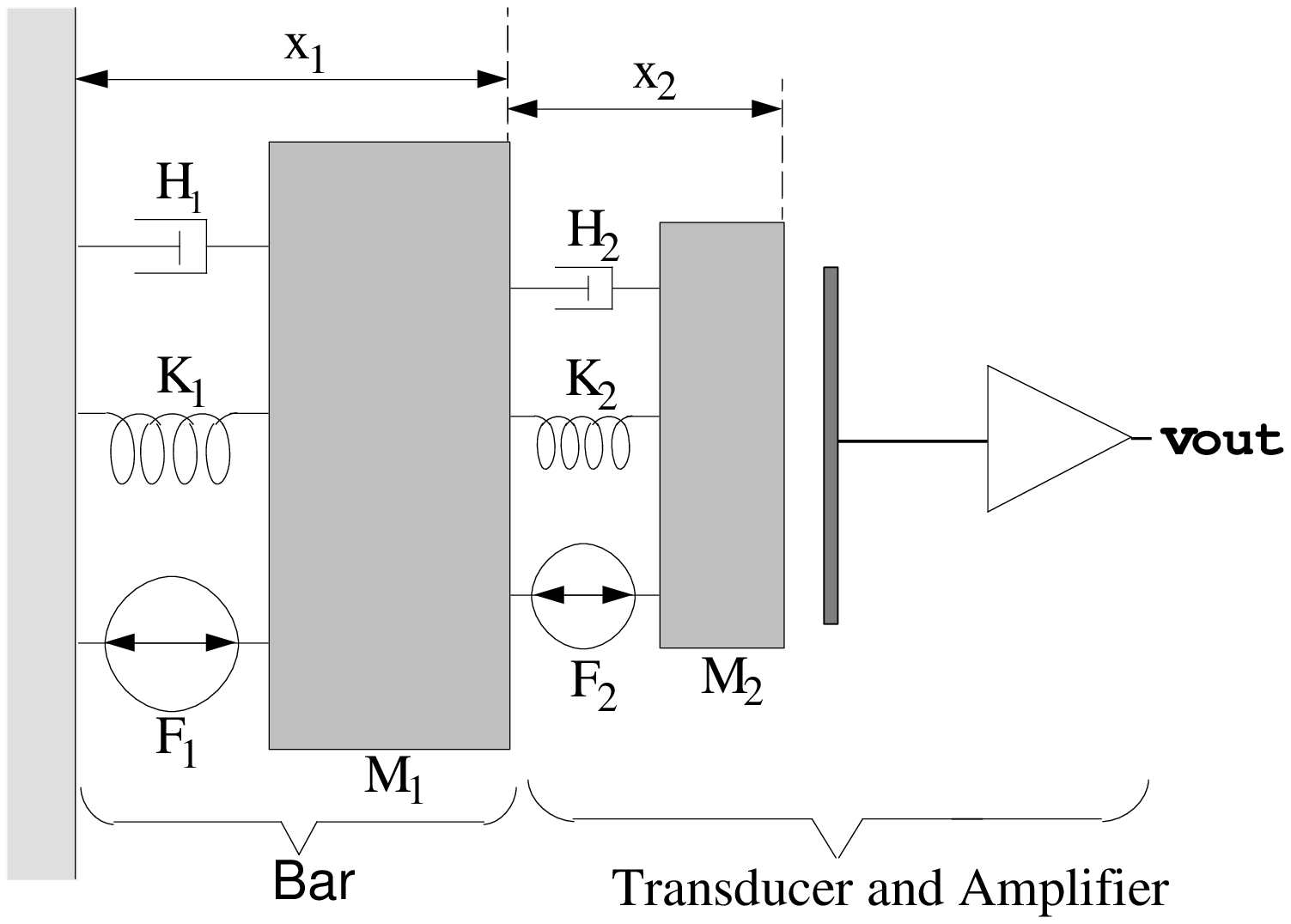 scaled 675}}
   \caption{A model of the Allegro antenna.}
   \label{modl}
\end{figure}

\begin{figure}
  	\centerline{\BoxedEPSF{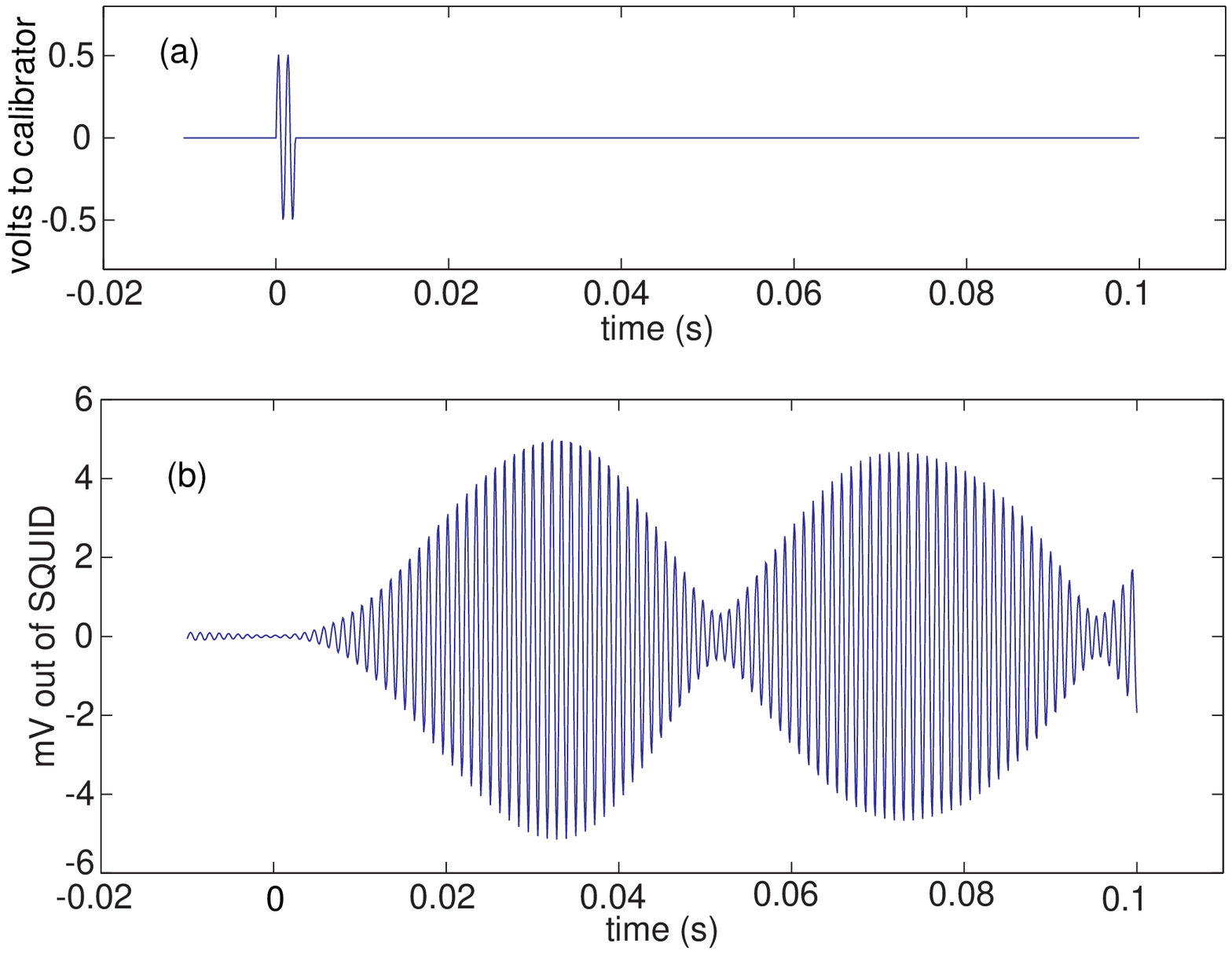 scaled 500}}
   \caption{Antenna response to a burst signal.  (a) The voltage provided to the
     calibrator.(b) The voltage response out of the SQUID in the time domain.} 
   \label{signl}
\end{figure}

The Allegro detector model is illustrated schematically in Fig.~\ref{modl}, where we
include all of the relevant stationary noise sources. The equations of motion for this
model are:

\begin{eqnarray}
     M_{1}\ddot{x}_{1}(t) + H_{1}\dot{x}_{1}(t) + K_{1}x_{1}(t)
   - H_{2}\dot{x}_{2}(t) - K_{2}x_{2}(t) \nonumber\\
   = F_{1}(t) - F_{2}(t) + F_{T}(t)
   + \frac{1}{2}M_{1}L_{1}\ddot{h}_{xx}(t) 
\label{force}
\end{eqnarray}
\begin{equation}
     M_{2}(\ddot{x}_{2}(t) + \ddot{x}_{1}(t)) + H_{2}\dot{x}_{2}(t) 
   + K_{2}x_{2}(t) = F_{2}(t) - F_{T}(t) 
\label{force2}
\end{equation}
$ M_{1,2} $ are the effective masses of the bar and mushroom resonator.
$ L_{1}$ is the effective length of the bar.  $K_{1,2}$ represent the spring constants of
the bar and mushroom resonator,  $H_{1,2}$ their respective damping coefficients. 
$F_{1,2}$ are the Langevin force noise generators associated with the dissipation
coefficients of each mass and
$F_{T}$ is the noise generated by a changing magnetic pressure from the superconducting
pick-up coil on the small mass resonator.  $x_{1}$ is the amplitude of the first
longitudinal normal mode of the bar, while 
$x_{2}$ is the relative displacement between the bar and the second resonator. The last
term on the right hand side of Eq.~(\ref{force}) is the component of the gravitational
wave tidal force along the bar axis. 
 The model shown does not explicitly include the superconducting circuitry or
the SQUID.  The voltage out of the SQUID is proportional to the relative displacement of
the two masses:
\begin{equation}
   V_{out}(t) = Gx_{2}(t) + \eta(t) 
\end{equation}   where G is a gain factor and $\eta(t)$ is white noise from the SQUID. 
The time response of the antenna to a large burst signal provided by the calibrator is
shown in Fig.~\ref{signl}.  The power spectrum of the stationary noise out of the SQUID
is shown in Fig.~\ref{resp}(a). The two resonant modes, seen clearly in the figure, are
at 896.8 Hz and 920.3 Hz.  We refer to them as the minus and plus modes respectively.
Figure~\ref{resp}(b) shows the antenna response to a large burst signal and (c) shows the
ratio of the noise to signal which is the stationary noise treated as if it were due to a
random flux of gravity waves exciting the bar.  
\begin{figure}
  	\centerline{\BoxedEPSF{94strain.eps scaled 750}}
   \caption{(a) The power spectrum of the voltage out of the SQUID due to stationary noise
     sources acting on the antenna. (b) The power spectrum after the bar has been excited
     by a large burst signal.  (c) The square root of the ratio of (a) and (b), showing the
     stationary noise as if it were all due to a random flux of gravity waves exciting the
     bar.  Also included in (c) is the predicted strain noise from the full Allegro model
    (dashed line).}
   \label{resp}
\end{figure}
\section{Data Acquisition} 
\label{sec:dataq}

\subsection{Signal Demodulation}

\begin{figure}
  	\centerline{\BoxedEPSF{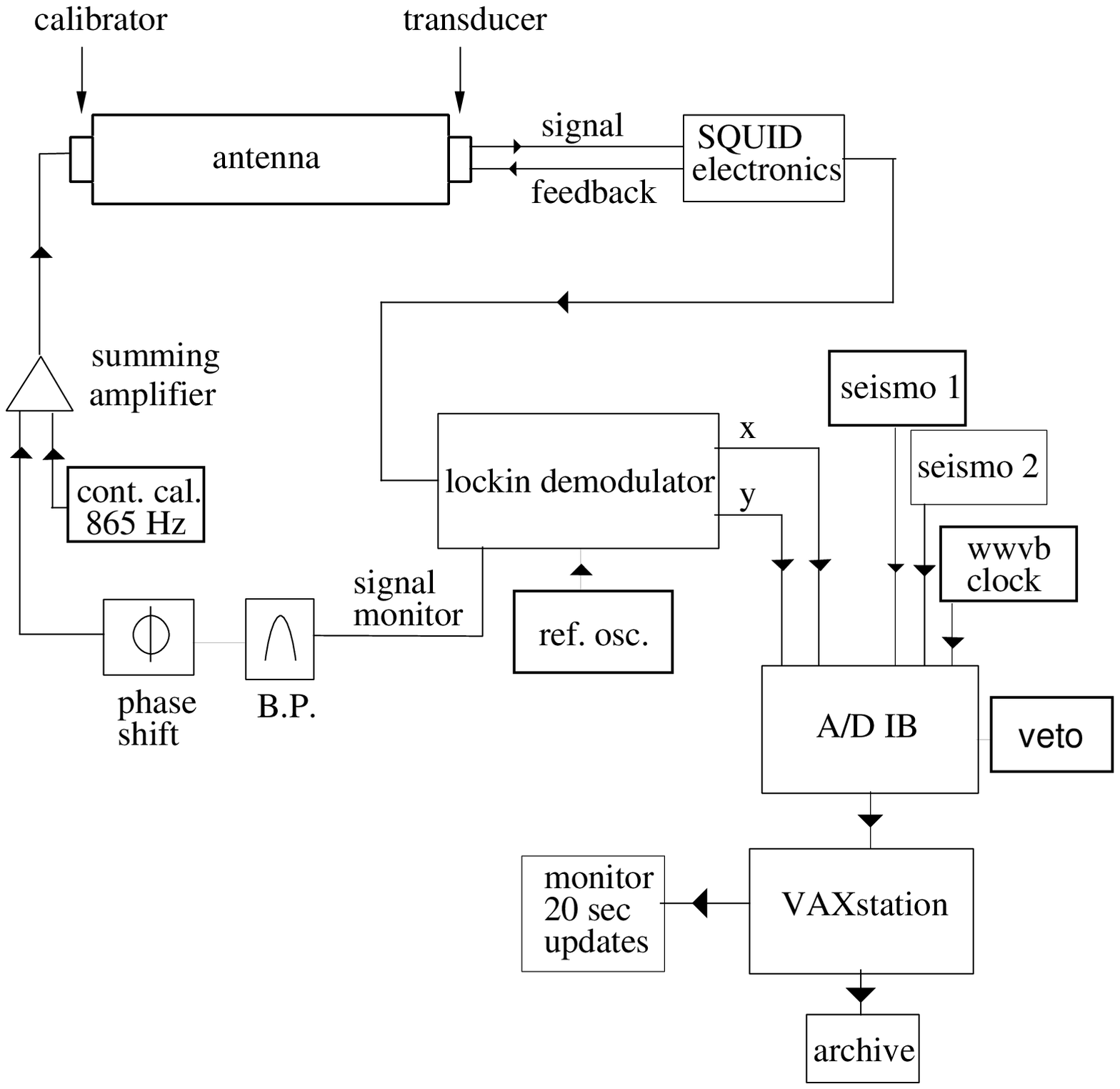 scaled 500}}
   \caption{Schematic of the Allegro data acquisition system.}
   \label{daqui}
\end{figure}

The voltage from the SQUID electronics is sent to a single lockin detector which
demodulates and low pass filters the signal.  The reference frequency of the lockin is
set halfway between the normal mode frequencies of the antenna, thus shifting the
frequency of the signal from the normal modes of the antenna to low frequency.  Because
the lockin is set for a wide bandwidth, the frequency response of the detector over its
entire bandwidth is monitored, enabling us to measure both the amplitude and phase of
each of the resonant normal modes.  It is due to the wide bandwidth that the continuous
systems test can be applied to the antenna at a frequency far enough removed from the
resonant modes as to not interfere with them. Other data collected to help monitor the
detector includes a direct low frequency signal from the SQUID, the status of the
hardware and SQUID vetos and signals from two seismometers, one of which is located on
the floor next to the dewar containing the antenna, the other on top of the vibration
isolation table.  A schematic of the data acquisition system in shown in Fig.~\ref{daqui}.
 The lockin is an EG\verb+&+G PAR 5210 two phase lockin amplifier with
reference frequency set at 908.5220 Hz, although this changes if the mode frequencies
shift by a couple of mHz.  The reference signal is provided to the lockin by a Hewlett
Packard 3325A function generator equipped with a high stability oscillator.  After the
lockin, the in-phase and quadrature output voltages are sent through an anti-aliasing
filter (cutoff frequency 40 Hz) and then to the A/D Interface Box (A/D IB). 

\subsection{Data Collection}

The A/D IB was constructed by the LSU Physics electronics shop and the Gravity Wave
Group.  It controls the sampling rate of the data, converts the analog signals to digital
data streams and sends the data to a VAXstation 3500.  An important factor to note is
that before any data was collected by the A/D IB rigorous testing was performed on it.  A
sine wave of amplitude large enough to span the entire range of the A/D converters was
input to each A/D and the output inspected to verify: (1) that the A/D converters
responded properly, (2) that the time between samples remained constant and that no
samples were missed, (3) that the data written to disk was consistent with the input
signal.  It was not until the data acquisition system ran for about a week without any
problems that it was considered stable enough to collect data.

A Kinemetrics model 60-DC clock provides a 1kHz square wave phase-locked to UTC which the
A/D IB uses as a counter.  When the counter reaches 8 ms the data is sampled and sent to
disk, resulting in a sampling time of 8 ms. Data sampled includes: (1) The in-phase and
quadrature output channels of the lockin (referred to as $x$ and $y$ respectively), (2) a
direct low frequency signal from the SQUID, (3) the signal from the two seismometers, (4)
the status of the hardware and SQUID vetos and (5) the sample time in UTC.  The voltages
out of the lockin are sampled with 16-bit accuracy, the others 12-bit accuracy.  This
data is referred to as the raw data to denote it has not been subject to any software
fiddling.  Table~\ref{rawtabl} shows the format of the raw data in a record.
\begin{table}[p]
\caption{Format of raw data record} 
\label{rawtabl}
\begin{tabular}{rl} \\
\multicolumn{2}{l}{once a record} \\ 40 bytes & Header \\ 
\multicolumn{2}{l}{2500 samples in a record of each of the following:} \\
 2 bytes & lockin output $x$  \\
 2 bytes & lockin output $y$  \\ 2 bytes &  2 seismometer outputs (12 bit resolution
each) \\ 2 bytes & 4 veto bits and 12 bits low frequency SQUID output  \\ 2 bytes &
universal time (unit seconds + milliseconds)(BCD encoded) \\ 
\multicolumn{2}{l}{in the header} \\ 2 bytes & UT day number \\ 1 byte & File identifier
(A...Z) \\ 1 byte & block type \\ 2 bytes & record number in file (1...4320) \\ 2 bytes &
run number \\ 8 bytes & VMS time of the first sample in the block \\ 6 bytes & spare \\ 4
bytes & university id \\ 2 bytes & universal time of the first sample in the block \\ 2
bytes & UT word 1 \\ 2 bytes & UT word 0 \\  2 bytes & gain code \\ 2 bytes & sampling
time \\ 2 bytes & number of samples in a block \\ 2 bytes & number of lockins \\
\end{tabular}
\end{table} Twenty seconds worth of data (2500 samples) is assembled by software into a
data block and written to disk.  There are 4320 blocks in a full day's worth of data.  A
DEC 3000 AXP, clustered to the VAXstation 3500, is used for on-line monitoring of the
detector and analysis of the data.  A week's worth of data (about 875 Mbytes) is allowed
to accumulate on disk and is then archived to 4mm DAT tape.  Two tapes are made using VMS
BACKUP to insure the data is transferred accurately and one tape is made using VMS COPY
which allows easy access to the data.  One BACKUP tape and the COPY tape are stored in
the lab, while the other BACKUP tape is stored off campus.

\section{Data Analysis} 
\label{sec:datanaly}

The data analysis programs read a block of data directly off of the disk.  Since the data
is written to disk every twenty seconds, it is available twenty seconds or less after
being collected.  In this way problems with the detector or with the electronics can be
identified and corrected quickly.  This ``on-line'' monitoring capability does not affect
the taking of data since it is a separate program running at a much lower priority than
the collection routine.

A single program, written in the language MATLAB\footnote{The MathWorks, Inc. 24 Prime
Park Way, Natick, Mass.01760}, does the majority of the data analysis.  The analysis
begins by reading in a block of data, removing DC offsets from the in-phase and
quadrature signal components and correcting for lockin gain. The program implements two
digital lockins which mix $x$ and $y$ with reference frequencies set at the plus and
minus resonant frequencies. The outputs of these digital lockins are the in-phase and
quadrature components of the amplitude of each mode, written as $ x_{\scriptscriptstyle
+}, x_{\scriptscriptstyle -}, y_{\scriptscriptstyle +}, y_{\scriptscriptstyle -}
$  where
$+$ refers to the plus mode and $-$ the minus.  The in-phase and quadrature components of
each mode are then separately filtered with an $8^{th}$ order digital Bessel
anti-aliasing filter having a corner frequency of 2.35 Hz.  The filtered data is then
decimated to reduce the amount of data handling. We keep only every tenth sample,
truncating the data to an effective sampling time of 80 ms. 

\begin{figure}[t]
  	\centerline{\BoxedEPSF{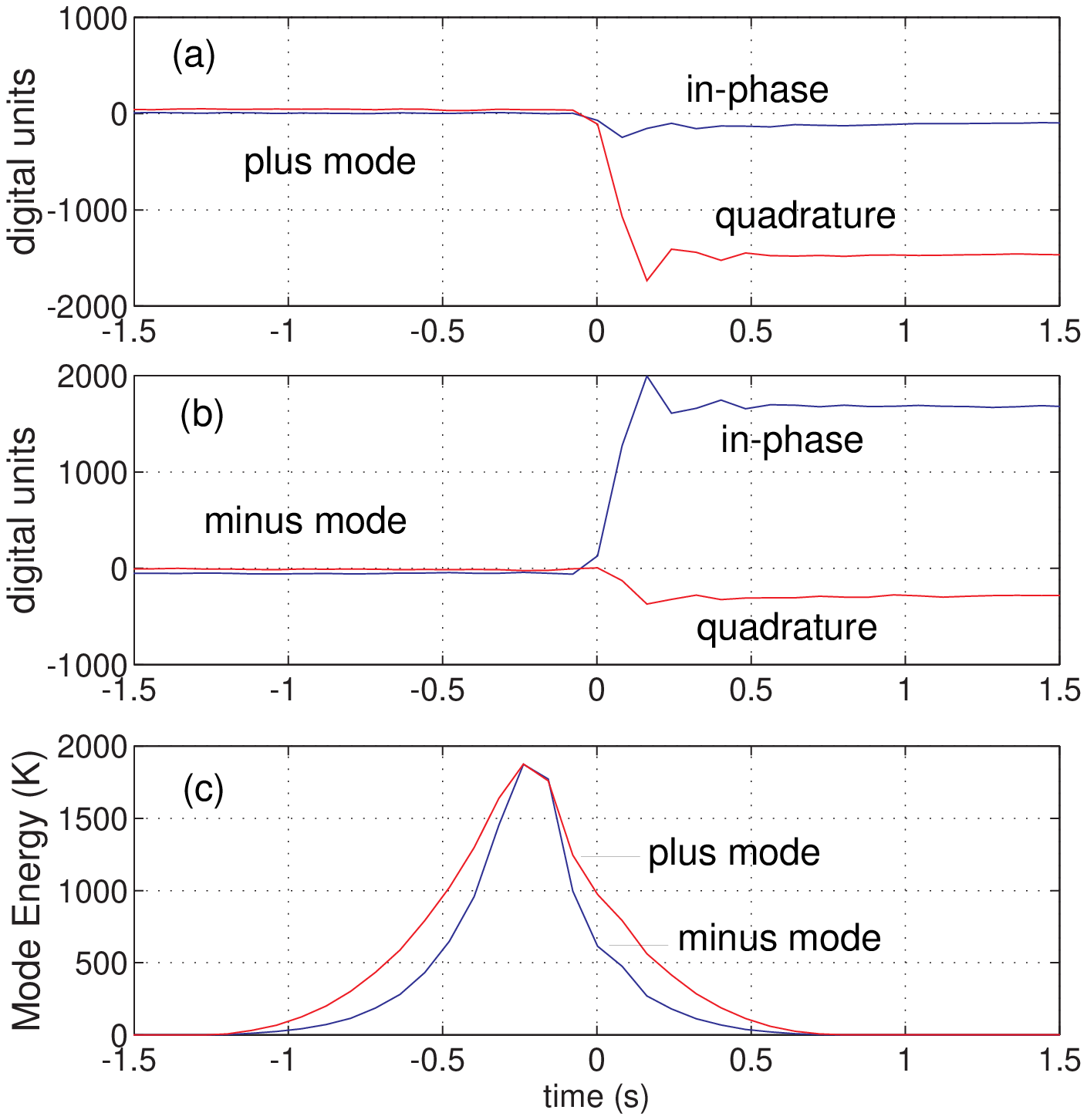 scaled 675}}
   \caption{(a) The plus and (b) the minus mode amplitudes in response to the signal of 
      Fig.~\protect\ref{signl}. (c) The optimally filtered response to the same signal.}
   \label{ben}
\end{figure}

The in-phase and quadrature components of each mode are optimally filtered for a burst
signal (see Sec.~\ref{sec:optfilt}) and the output squared and added to form the mode
burst energies at each sample.  Representing the output of the optimal filter by 
$f_{x \scriptscriptstyle +}, f_{ y \scriptscriptstyle +},$
$f_{x \scriptscriptstyle -}$ and $f_{y \scriptscriptstyle -}$, the estimate of the burst
energy at each sample is
\begin{equation}
 E_{\scriptscriptstyle \pm} = f_{x \scriptscriptstyle \pm}^{2} + f_{y \scriptscriptstyle
\pm}^{2}.
\label{moden}
\end{equation} The mode response to a large burst both before and after optimal filtering
is shown in  Fig.~\ref{ben}.
 It has become conventional in this field to express energy in Kelvin. 
Therefore, a ``mode noise temperature'' is defined as the mean value of the mode burst
energy
$T_{\scriptscriptstyle \pm} = <E_{\scriptscriptstyle \pm}>/k_{B}$.  Burst energy is not
to be confused with the energy in a mode as given by the equipartition theory. Instead,
it is a measure of the change in energy of the modes between samples.  Since the sampling
time is much less than the ``random walk'' time of the antenna (8 ms compared to 40 mins)
the noise temperature is much less than the physical temperature of 4.2 K.

\begin{figure}
  	\centerline{\BoxedEPSF{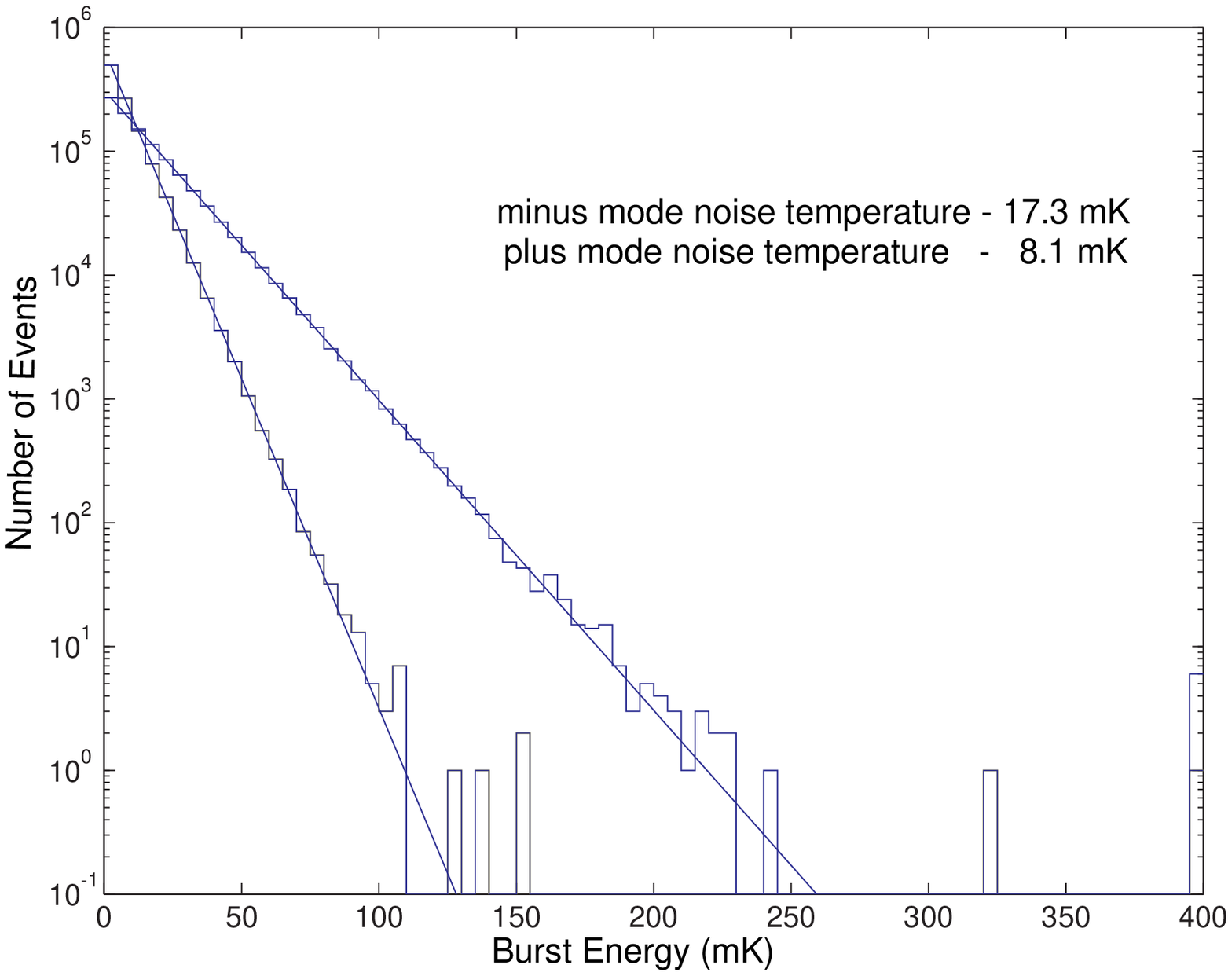 scaled 500}}
   \caption{ A histogram of the energy in each mode for day 200 of 1994.  The
     slope of the histogram gives the noise temperature for each mode.}
   \label{hist}
\end{figure}

To reduce the amount of data handled, a threshold is applied  so that only those samples
with energy $10\times$ the noise temperature or greater in both modes are recorded each
day by the analysis programs.  Each sample is tagged with the time in seconds from the
start of the day.  Above this threshold there are roughly 400-600 Allegro samples per day
(Fig.~\ref{hist}).   Also as part of the analysis the average over each record of
$ x_{\scriptscriptstyle +}, x_{\scriptscriptstyle -}, y_{\scriptscriptstyle +},
y_{\scriptscriptstyle -}$, $E_{\scriptscriptstyle +}$ and 
$E_{\scriptscriptstyle -}$ is recorded along with the UTC time of the start of each
record and the raw low frequency and seismometer data.  This information is used
primarily for diagnostic checks on the detector. 

\begin{figure}
  	\centerline{\BoxedEPSF{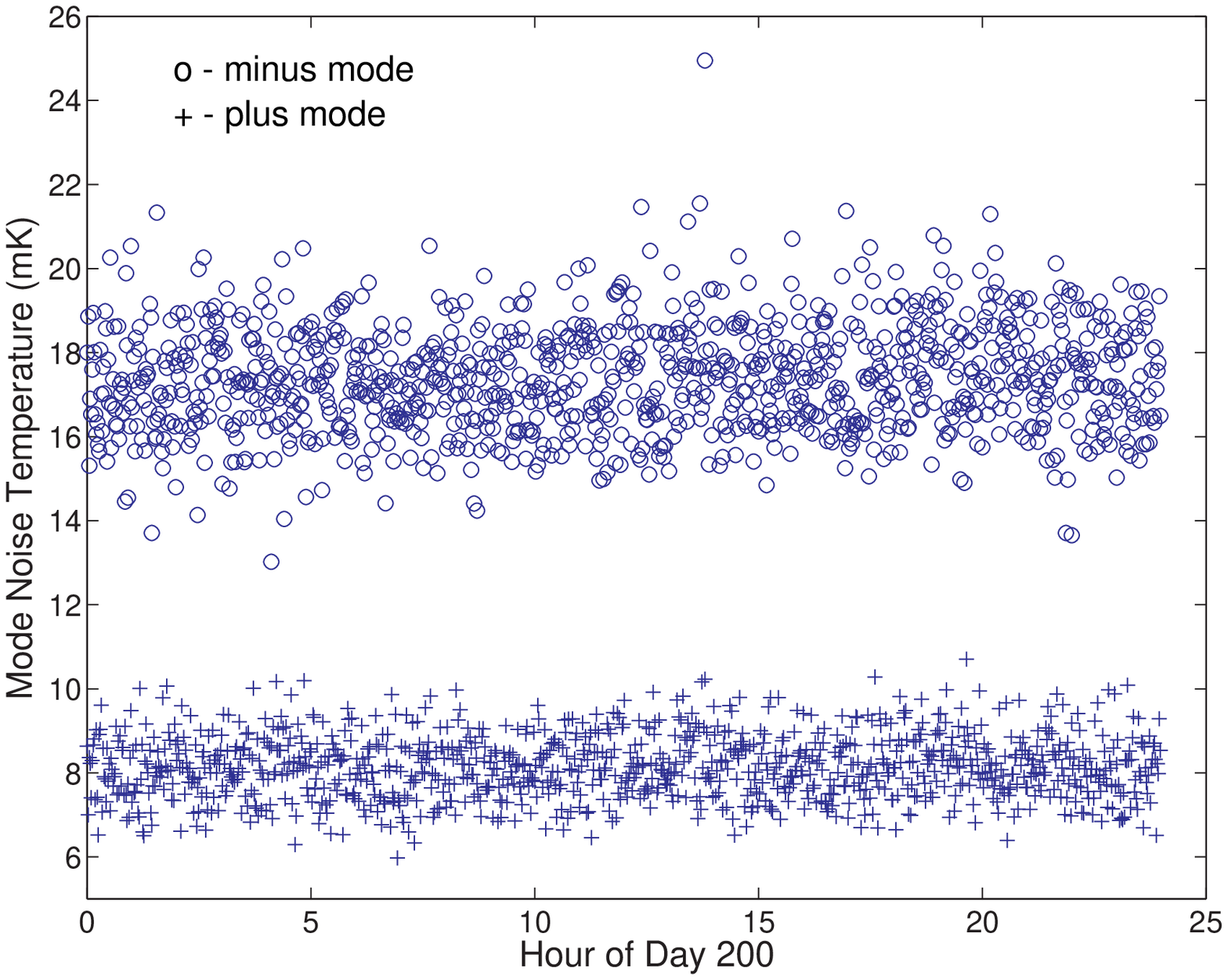 scaled 500}}
   \caption{The average noise temperature in each mode during day 200 of 1994.  Each data
point
   represents a six minute average of the mode burst energies.}
   \label{Ntemp}
\end{figure}

\begin{figure}
  	\centerline{\BoxedEPSF{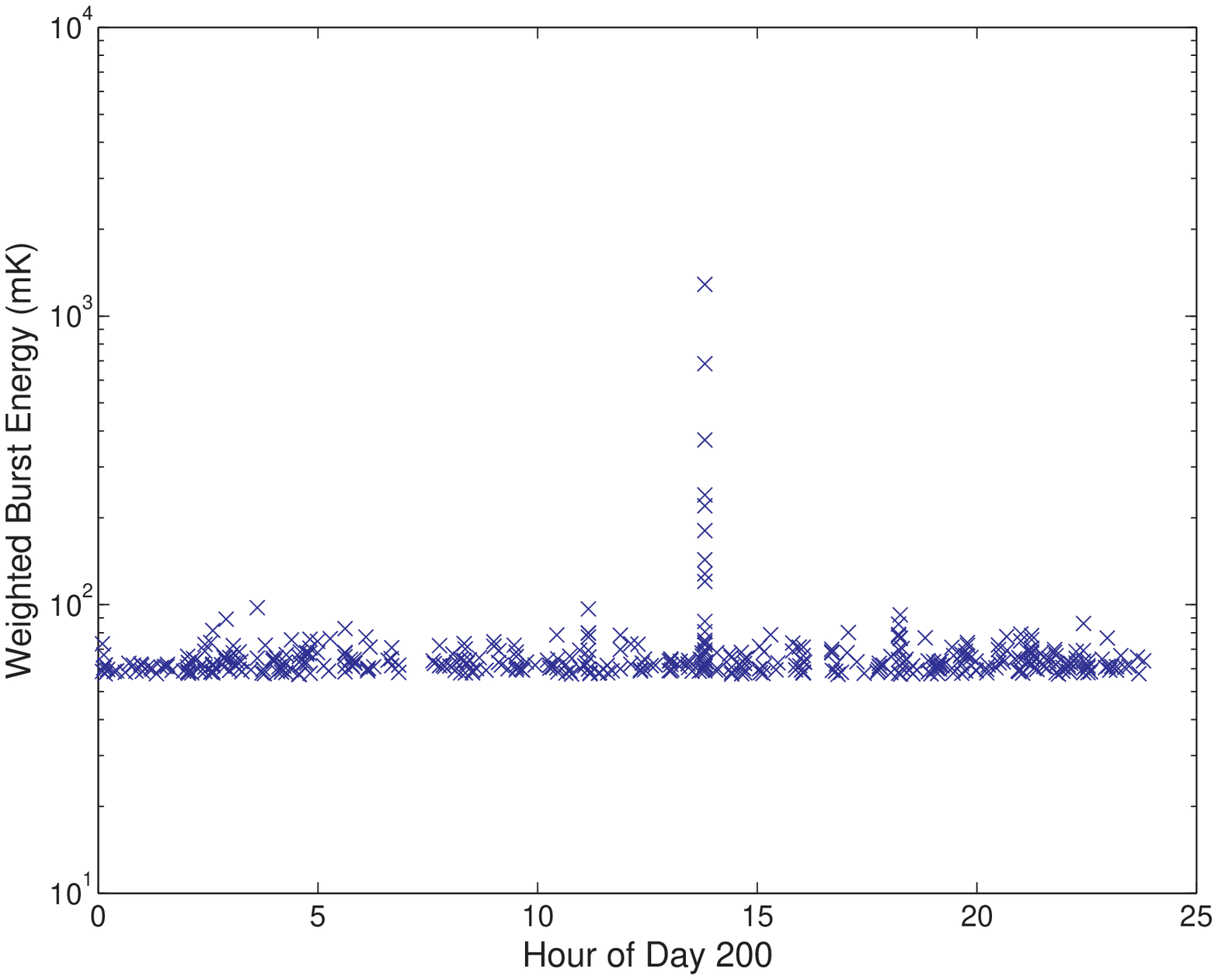 scaled 500}}
   \caption{The final event list for day 200 of 1994.}
   \label{elist}
\end{figure}

This is the end of the analysis unless there is to be a coincidence search with other
gravity wave detectors, such as the Explorer detector of the University of Rome, or GRO
gamma ray data. Before exchanging data, we first edit those excitations of the antenna
that can be positively identified as something other than a gravity wave (such as an
earthquake or an electronic hiccup).  Next the mode noise temperatures (Fig.~\ref{Ntemp})
are calculated in six minute averages for the entire span of the coincidence search.  The
statistically correct way to combine the energy information from both modes is by forming
a weighted burst energy 
\begin{equation}
 E_{w} = T_{w}(E_{\scriptscriptstyle +}/T_{\scriptscriptstyle +} + 
 E_{\scriptscriptstyle -}/T_{\scriptscriptstyle -}) 
\label{wbe}
\end{equation} where 
\begin{equation}  T_{w}^{-1} = T_{\scriptscriptstyle +}^{-1} +  T_{\scriptscriptstyle
-}^{-1} 
\end{equation} is the weighted noise temperature (this is the overall noise temperature
of the detector) and $T_{\scriptscriptstyle +}, T_{\scriptscriptstyle -}$ are the
previously mentioned averages.
  A threshold is applied to $E_{w}$ so that only samples with
$E_{w}>11.5T_{w}$ are kept.  The factor of 11.5 was chosen so that the Allegro event rate
for the 1991 coincidence search with the Explorer detector would be about 100
events/day.  The consistency of the Allegro detector is demonstrated by the fact that the
same threshold produced about 100 events/day for the entire 3 1/2 years of continuous
operation. Consecutive samples above threshold are then collapsed into a single time and
energy, creating an event.  The energy assigned to the event is the energy of the sample
in the series of consecutive samples above threshold with the maximum energy value.  The
time of the event is given by the time of the first sample in the series plus half the
duration of the series, where the duration is defined to be the time of the last sample
minus the time of the first sample.  The sample time is determined by reading the UTC
time at the beginning of the record containing the event and then counting the number of
samples (at 80 ms between samples) into the record to that event.  Then an offset is
subtracted from the resulting time to account for filtering delays. Figure~\ref{elist}
shows the final event list for a small section of data.

\section{The Filtering Algorithm} 
\label{sec:optfilt}

The optimal filtering on Allegro is done in the time domain so that it can be applied
directly to the incoming data.  We use the MATLAB {\em filter} routine which applies the
optimal filter to the data using the transposed direct form II
structure~\cite[p.155]{oppenheim_psd75}.  The filter coefficients which will maximize the
signal to noise ratio for a sequence of data involving stationary noise are given by 
~\cite[pages 184--184]{wan_zub}, 
\cite[pages 126--135]{whalen}
\begin{equation}
 \underline{a} = \underline{\underline{R}}^{-1}\underline{s} 
\label{wz}
\end{equation}   where $\underline{a} $ is the vector containing the filter weights,
$\underline{\underline{R}}^{-1}$ is the inverse of the autocorrelation matrix of the
noise and
$\underline{s}$ is the detector's response to the signal being looked for. In the
following analysis of the optimal filter we will use a single underline to denote a
vector and two underlines for a matrix.  Because the in-phase and quadrature components
for each mode are statistically similar, they can be averaged to make a single
correlation function for the noise in each mode.  Also, the in-phase and quadrature
components of the signal vector are combined (described later) to form the mode response
to a burst.  The correlation functions of the two modes are not similar and therefore a
pair of filter weights are created, one to filter the plus mode and the other to filter
the minus mode.  The details of creating the autocorrelation matrix and the response
vector (signal) are described next. 

\subsection{The Signal}

\begin{figure}[p]
  	\centerline{\BoxedEPSF{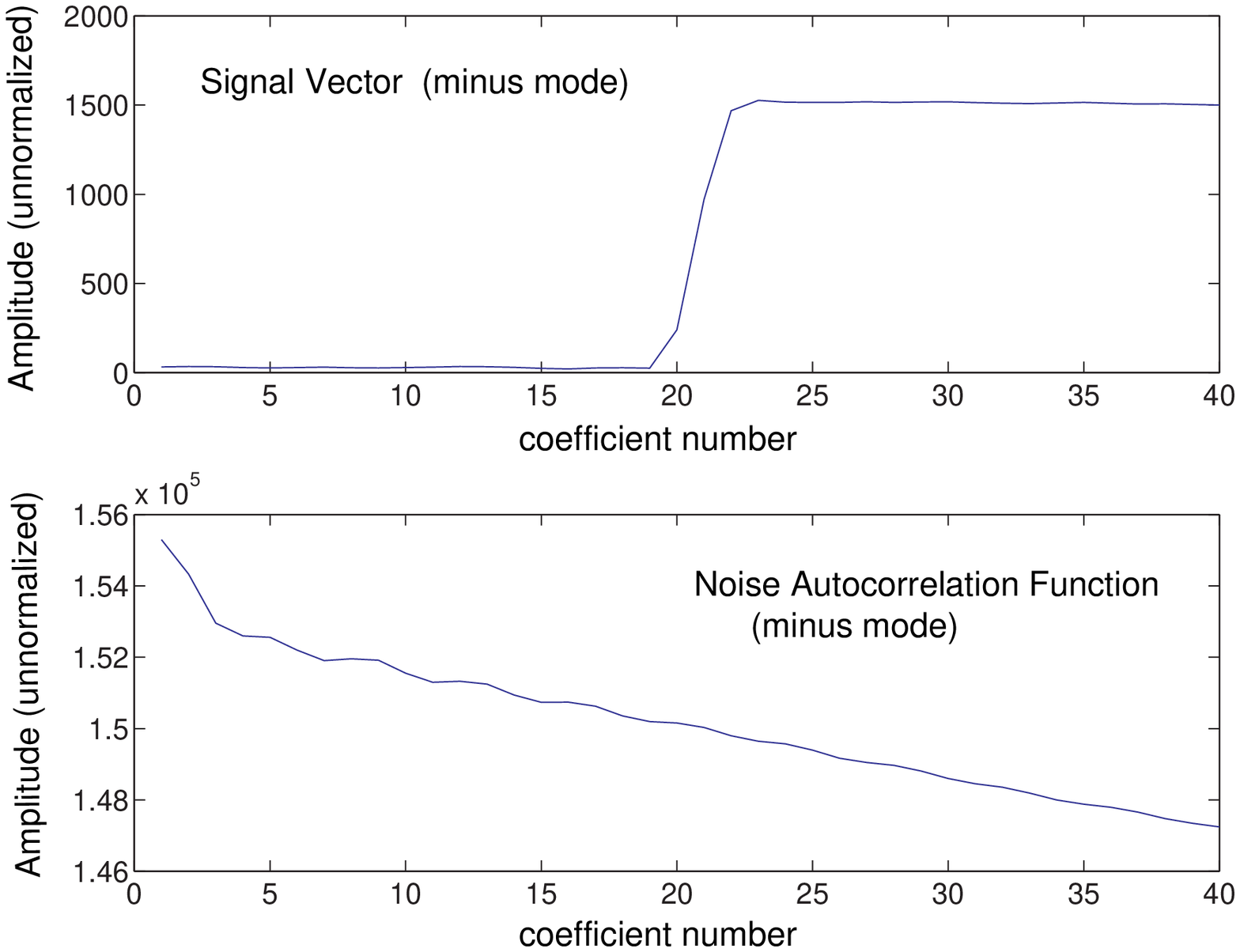 scaled 500}}
   \caption{The signal vector and noise autocorrelation function used in creating the
optimal
   filter for the minus mode.}
   \label{Nsig}
\end{figure}

The signal vector $\underline{s}$ is obtained by applying a very large calibration pulse
to the antenna so that the low pass filtered and decimated output is essentially
unaffected by the stationary noise.  Next, the mean value of the first few samples is
subtracted from each sample in the signal array so that the amplitude just before the
pulse hits the antenna is near zero.  Then, the squares of the in-phase and quadrature
signal components in each mode are added and the square root taken to create the final
form of the signal vector (Fig.~\ref{Nsig})
\begin{equation}
 \underline{s_{\scriptscriptstyle \pm}} = \sqrt{\left( \underline{s_{\scriptscriptstyle
\pm}^{x}} \right)^{2} + \left( \underline{s_{\scriptscriptstyle \pm}^{y}} \right)^{2}}. 
\end{equation} 

\subsection{The Noise}

The first step in forming the autocorrelation matrix for the noise is to form the
autocorrelation function for one record's worth of low pass filtered and decimated data 
\begin{equation}
 \underline{R} = \frac{1}{N} \sum_{i=0}^{N-1} n_{i}n_{i+j} 
\end{equation} with $N$ the number of coefficients in the filter, $i$ the sample index
and $j$ the time offset index.  This is done every twentieth record for an entire day's
worth of data.  It is necessary to use such a long time span of data because of the long
relaxation times of the normal modes. All events outside the thermal distribution are
removed from the data before forming the correlation function as the presence of
non-stationary noise will degrade the filter's performance.  The length of the filter,
$N$, was determined experimentally.  Filter lengths of 20 to 50 decimated samples were
tried and it was found that the noise temperature of the modes decreased up to 40
coefficients (amounting to 3.2 seconds of data). After that, the noise temperature no
longer decreased with increasing coefficient number so 40 decimated samples was chosen as
the length for the filter.  

\begin{figure}[p]
  	\centerline{\BoxedEPSF{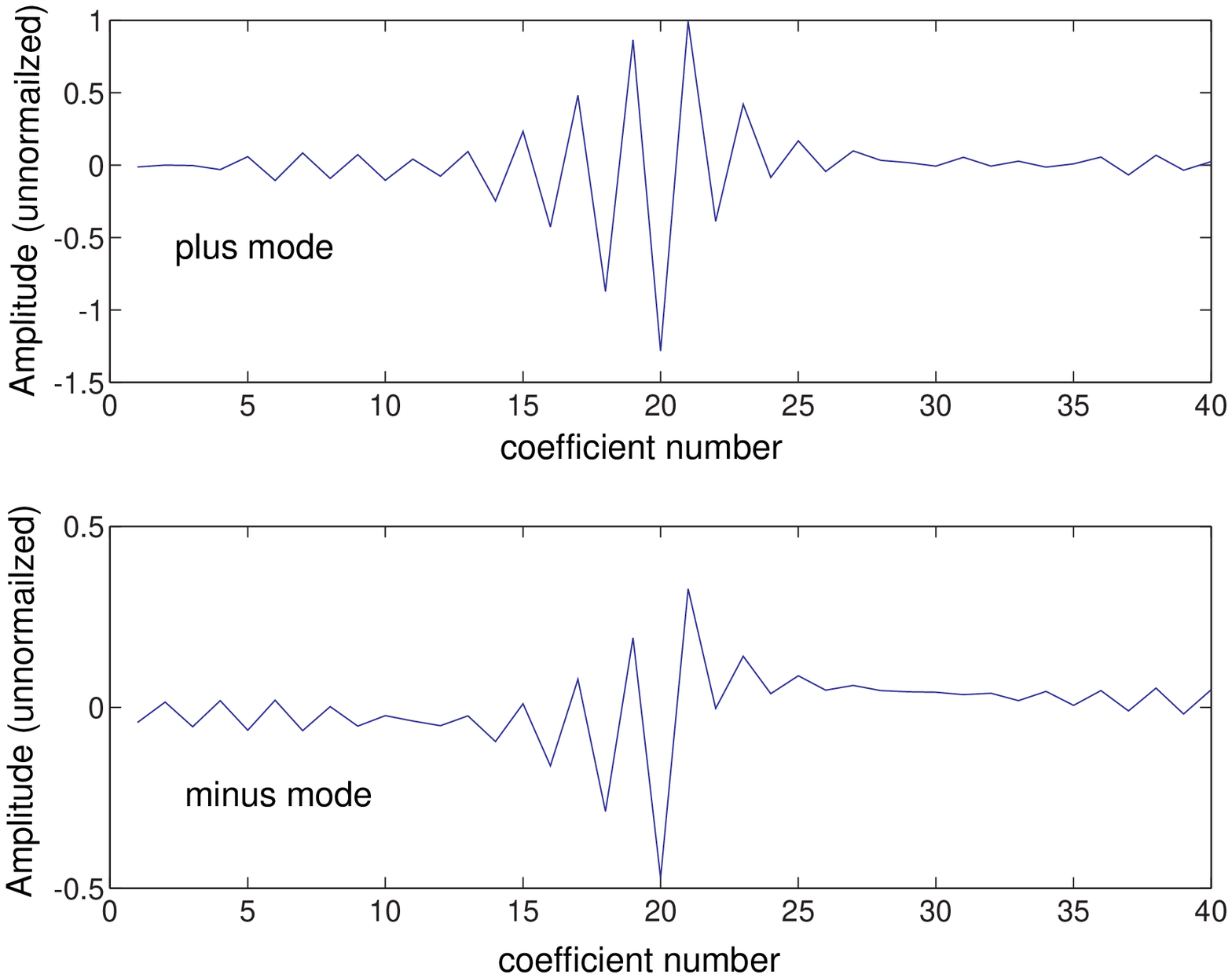 scaled 500}}
   \caption{The optimal filter weights for the minus and plus modes.}
   \label{optfil}
\end{figure}

Next, the values of $\underline{R}$ at each $j$ from each record analyzed are summed and
the in-phase and quadrature components added to form the autocorrelation function for a
mode.  The autocorrelation matrix is formed using the Matlab routine {\em toeplitz} such
that the zero delay components $R_{00}$ lie along the diagonal
\begin{equation}
 \underline{\underline{R_{\scriptscriptstyle \pm}}} = toeplitz(\frac{1}{2}
(\underline{R}^{x}_{\scriptscriptstyle \pm} + \underline{R}^{y}_{\scriptscriptstyle
\pm})). 
\end{equation}
 Here the in-phase and quadrature components are denoted with an $x$ and $y$
respectively.  The inverse of the matrix is formed using the Matlab {\em inv} routine
\begin{equation}
  \underline{\underline{R_{\scriptscriptstyle \pm}}}^{-1} = inv
  \left(\underline{\underline{R_{\scriptscriptstyle \pm}}} \right). 
\end{equation} 
 Having obtained the inverse of the autocorrelation function for the noise
and the signal vector, the filter weights for both the plus and minus modes are formed by
Eq.~(\ref{wz}).  The final form of the weights is shown in Fig.~\ref{optfil}. 

\subsection{Normalization}

Once the optimal filter is constructed the weights are normalized by putting a pulse of
known energy into the antenna using the calibrator.  A SRS Model DS345 function generator
was used to provide 2 cycles of a 908 Hz sine wave of constant amplitude to the
calibrator at twenty second intervals.  The energy deposited in each mode of the antenna
by a single pulse is~\cite{calb}
\begin{equation}
  E_{cal \scriptscriptstyle \pm} = \frac{\pi}{2}
  \frac{\gamma_{\scriptscriptstyle \pm}}{\omega_{\scriptscriptstyle \pm}}(NV)^{2}
\label{cal}
\end{equation}  where $\gamma_{\scriptscriptstyle \pm}$ is the calibrator coupling
coefficient for each mode, defined as the ratio of the current output from the calibrator
to the input driving voltage and were measured in a separate experiment. 
$\omega_{\scriptscriptstyle \pm}$ are the mode frequencies, N is the number of cycles and
V is the voltage zero to peak provided to the calibrator.  A weighted burst energy is
formed with Eq.~\ref{wbe}. 

Approximately 60 pulses were applied having an amplitude large enough so that the effects
of stationary noise on the estimate of the resulting burst energies was small.  After
applying the pulses the data was analyzed as described in Section~\ref{sec:datanaly} to
produce a list of corresponding events.  The mean of the mode burst energies of the
pulses was compared to the expected energy deposited in the antenna as given by
Eq.(\ref{cal}) and the filter weights scaled so that the two matched.  

The filtering scheme causes a delay between the actual arrival time of a pulse and the
recorded arrival time.  This delay needs to be measured and removed from the estimate of
the timing of events. A very large calibration pulse was applied to the antenna and
analyzed to produce a filtered event.  This event consisted of approximately 40 decimated
samples similar to Fig.~\ref{ben}.  The time ascribed to this event by the procedure
described previously was 14581.672 seconds.  Examining the raw data (after lockin and low
pass filtering but before any processing by the analysis programs or decimation) it was
determined that the first signs of the calibration pulse effecting the antenna appeared
at 14579.936 seconds.  Subtracting the two gives a delay of 1.74 seconds, which is then
removed when the event times are recorded. 

\section{Event Uncertainties} 

It is impossible for a single detector to differentiate between a gravity wave passing
through the antenna and excitations due to noise.  At low energies the thermal spectrum
(stationary noise) masks any signal, while above that a signal is indistinguishable from
a burst of non-stationary noise.  Two or more (the more the better) detectors operating
in coincidence, however, can greatly reduce the noise level by demanding that: (1) a
gravity wave excite each antenna simultaneously within a few milliseconds, depending on
the distance between them and (2) for similar detectors aligned with respect to
astrophysical sources, such as Allegro and Explorer, the energy deposited in each be
equal.  Unfortunately, noise sources add a degree of uncertainty to any measurements of
event arrival time and energy with the result that one looks instead for a coincidence:
(1) in a window of time which is much greater than the light travel time between
detectors and (2) where the energy of a signal is no longer equal in each detector, but
lies in some range which we shall show depends on  both the noise temperature of the
detectors and the strength of the signal.  In this section we quantify the uncertainties
introduced into these measurements by the stationary noise.

Using the same parameters as described in Section~\ref{sec:optfilt} a series of
calibration pulses was applied to the bar.  A signal from the function generator was
connected to one channel of the hardware veto so that at the same time a voltage pulse
was applied to the calibrator a veto was recorded.  This allowed the sample immediately
following excitation of the antenna to be identified.  The largest error this procedure
can produce in the timing of the pulse is 8 ms which, as we shall see, is much smaller
than the final uncertainties in the timing.  The anticipated energy deposited by each
pulse given by Eq.(\ref{cal}) and their number for each calibration series are shown in
Table~\ref{caltabl}. \\
\begin{table}[p]
   \caption{The anticipated energy of each calibration pulse and the number of pulses
applied
   at that energy for a given series.}
   \label{caltabl}

   \begin{tabular}{lccc} \\
    & calibration series &  pulse energy (mK) &  number of pulses \\ 
   & cal2\_312 & 64 & 100 \\
		 & cal3\_312	& 76 &  100 \\
   & cal4\_312 & 110 &  100 \\   
   & cal1\_317 & 220 &  100 \\
			& cal2\_317 & 420 &  100 \\
   & cal3\_317 & 1000 &  100 \\  
   & cal1\_327 & 110 &  60 \\
			& cal2\_327 & 150 &  60 \\
   & cal3\_327 & 220 &  60 \\  

   \end{tabular}
\end{table}

\subsection{Uncertainties in Timing}

\begin{figure}
  	\centerline{\BoxedEPSF{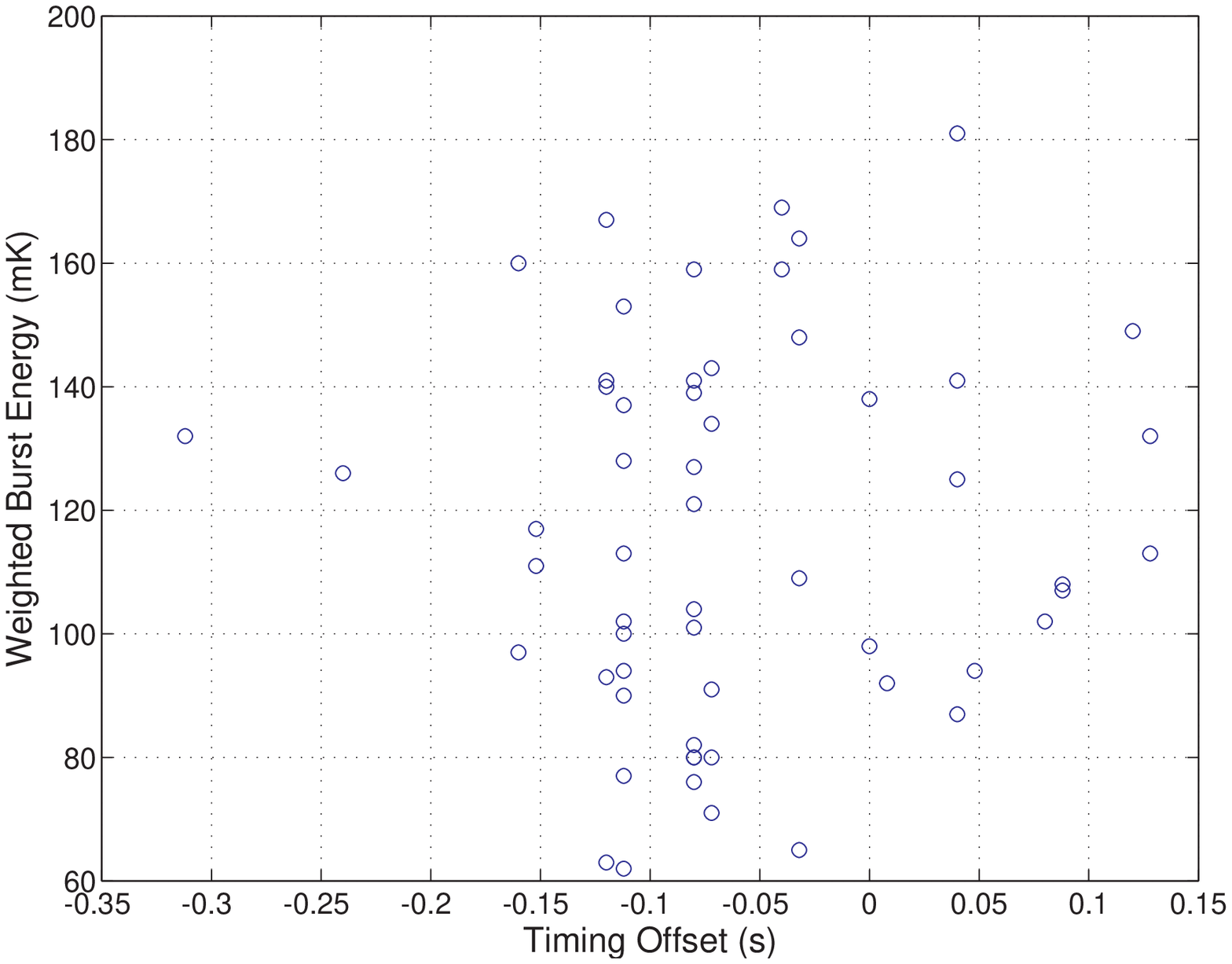 scaled 500}}
   \caption{The burst energy associated with each calibration pulse in series
   cal1\_327 and the corresponding timing offset.  The shift away from
   zero delay is common to all calibration series having a mean value of
   $\leq .06 $ s.}
   \label{offset}
\end{figure}

\begin{figure}
  	\centerline{\BoxedEPSF{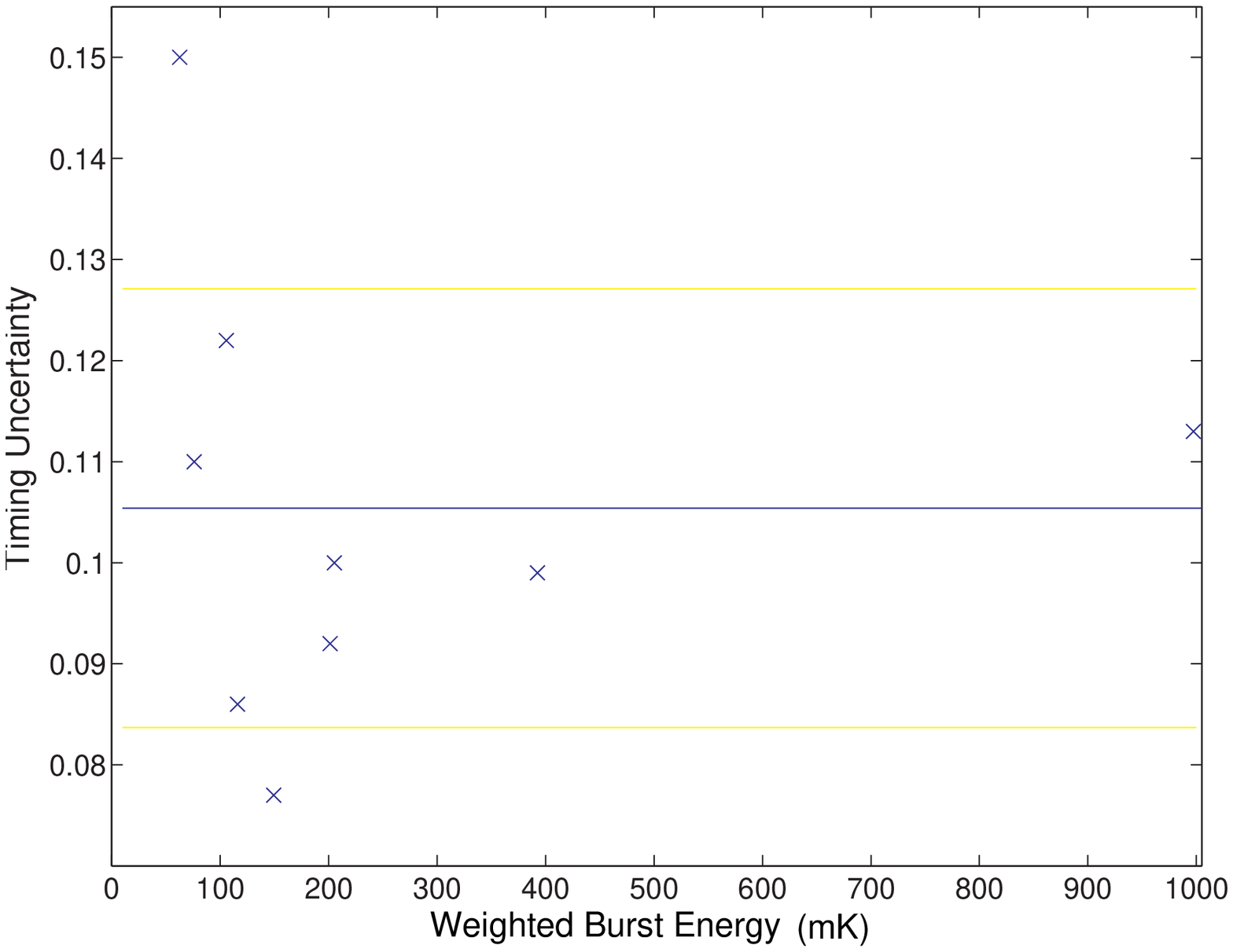 scaled 500}}
  \caption{ The uncertainty in the timing for each calibration series (defined  as the
standard
  deviation of the offsets) is represented by the  crosses.  The solid line in the mean
value
  of the timing uncertainties. The dotted lines are one standard deviation away from the
   mean value of the nine data points plotted).}
   \label{tunc}
\end{figure}

The calibration pulses were analyzed with the procedures described in Section
~\ref{sec:datanaly} to produce lists of event times and energies. Call the event time
assigned to each calibration pulse the ``arrival time'' and identify a ``pulse
application time'' with the tripping of the veto.  Subtracting the arrival time from the
application time produces a timing offset for each calibration pulse. Figure~\ref{offset}
shows the offset for each pulse in the calibration series cal1\_327.  The standard
deviation of the offsets is a measure of the uncertainties in our timing procedures. 
Figure~\ref{tunc} shows the standard deviation of the offsets from each series of pulses,
in effect plotting the timing uncertainty as a function of signal strength.  Of the nine
data points shown, 7 are within one standard deviation of the mean uncertainty, and only
the smallest signal (still about $11.5 T_{w}$) is significantly different.  What is
surprising is that at the highest signal strength, about
$180T_{w}$, the timing is no more accurate than at lower energies.  The end result is
that for signals above 
$11.5 T_{w}$ Allegro's timing is accurate to $\pm 0.1 $ second. 

\subsection{Uncertainties in Energy}

Noise sources acting on the antenna, both thermal and electronic, are stationary
distributed with zero mean.  This property is not changed by either the lockin or the
optimal filtering.  In the absence of a signal or non-stationary noise the outputs from
the optimal filter are statistically independent zero mean stationary variables with
variance
$\sigma_{\scriptscriptstyle \pm}^2$.  Forming the mode energies by Eq.~\ref{moden}
results in an exponential distribution  

\begin{equation}
 p(E_{\scriptscriptstyle \pm}) = \frac{1}{T_{\scriptscriptstyle
\pm}}\exp{\frac{E_{\scriptscriptstyle \pm}}{-T_{\scriptscriptstyle \pm}}} 
\end{equation}  with $E_{\scriptscriptstyle \pm}$ the detector response to stationary
noise and $T_{\scriptscriptstyle \pm}\equiv 2\sigma_{\scriptscriptstyle
\pm}^2$. This distribution has a non-zero mean given by

\begin{equation}
 <p(E_{\scriptscriptstyle \pm})> = T_{\scriptscriptstyle \pm}\ . 
\end{equation}

If a signal of burst energy $E_{cal\scriptscriptstyle \pm}$ is present, it can be shown
that the mode burst energy of the signal combined with the stationary noise is non-central
$\chi^{2}$ distributed with two degrees of freedom ~\cite{whalen}
\begin{equation}
  p(E_{\scriptscriptstyle \pm}) =
\frac{1}{T_{\scriptscriptstyle \pm}}\exp{\frac{(E_{\scriptscriptstyle
\pm}+E_{cal\scriptscriptstyle \pm})}{-T_{\scriptscriptstyle \pm}}} 
   I_{o}(\frac{2\sqrt{E_{\scriptscriptstyle \pm} E_{cal\scriptscriptstyle
\pm}}}{T_{\scriptscriptstyle \pm}}) 
\label{chi2}
\end{equation}   where
$I_{o}$ is the modified Bessel function of zeroth order. The mean and variance of this
distribution are given by

\begin{equation} 
 <p(E_{\scriptscriptstyle \pm})> \ = E_{cal\scriptscriptstyle \pm}+ T_{\scriptscriptstyle
\pm} 
\end{equation}
\begin{equation}
  var(p_{\scriptscriptstyle \pm}) = 2E_{cal\scriptscriptstyle \pm} T_{\scriptscriptstyle
\pm} + T_{\scriptscriptstyle \pm}^2  . 
\end{equation}

The weighted energy as defined in Eq.~(\ref{wbe}) is fourth order non-central
$\chi^{2}$  distributed ~\cite{whalen} with non-central parameter equal to the weighted
burst energy $E_{cal}$ 

\begin{equation}
  p(E_{w}) \ =
  \frac{1}{T_{w}}\exp{\frac{(E_{cal
  }+E_{w})}{-T_{w}}}I_{1}(\frac{2\sqrt{E_{w}E_{cal}}}{T_{w}})   
        \sqrt{\frac{E_{w}}{E_{cal}}} \  
\label{chi4}
\end{equation}  where $E_{w}$ is the weighted burst energy due to the stationary noise.
The mean and variance are  given by 
\begin{equation}
  <p(E_{w})> \ = E_{cal}+2T_{w} 
\end{equation} 
\begin{equation}
  var(p_{w}) = 2E_{cal}T_{w}+2T_{w}^2 \ . 
\label{energyvariance}
\end{equation}

\begin{figure}
  	\centerline{\BoxedEPSF{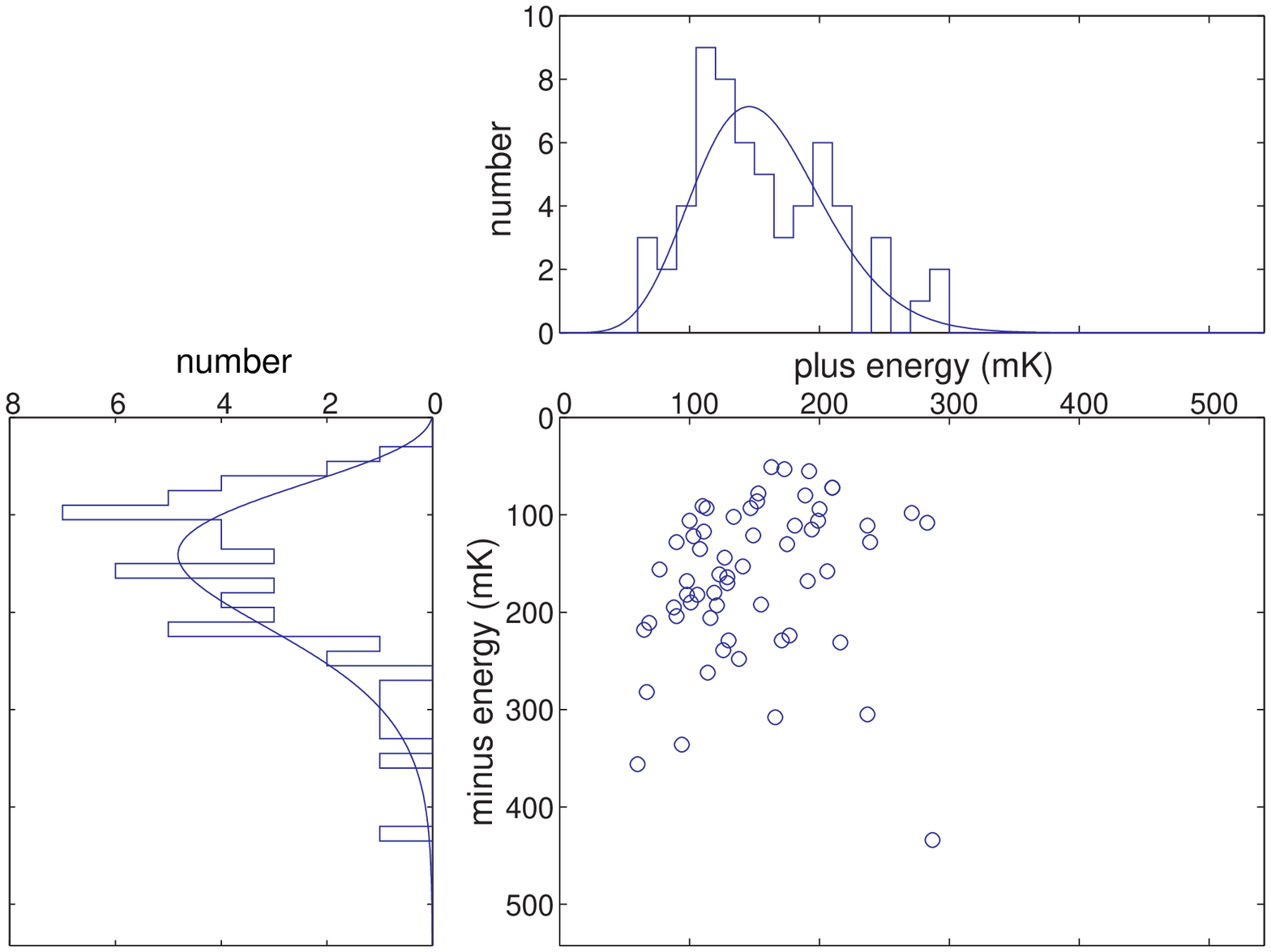 scaled 500}}
   \caption{The estimate of the energy deposited in the plus mode plotted against that in
the
   minus mode for each pulse in the calibration series cal2\_327 .  Sharing the same axis
are
   histograms of the mode burst energies and the distribution function of
   Eq.~(\protect\ref{chi2}).} 
   \label{chi}
\end{figure}

\begin{figure}
  	\centerline{\BoxedEPSF{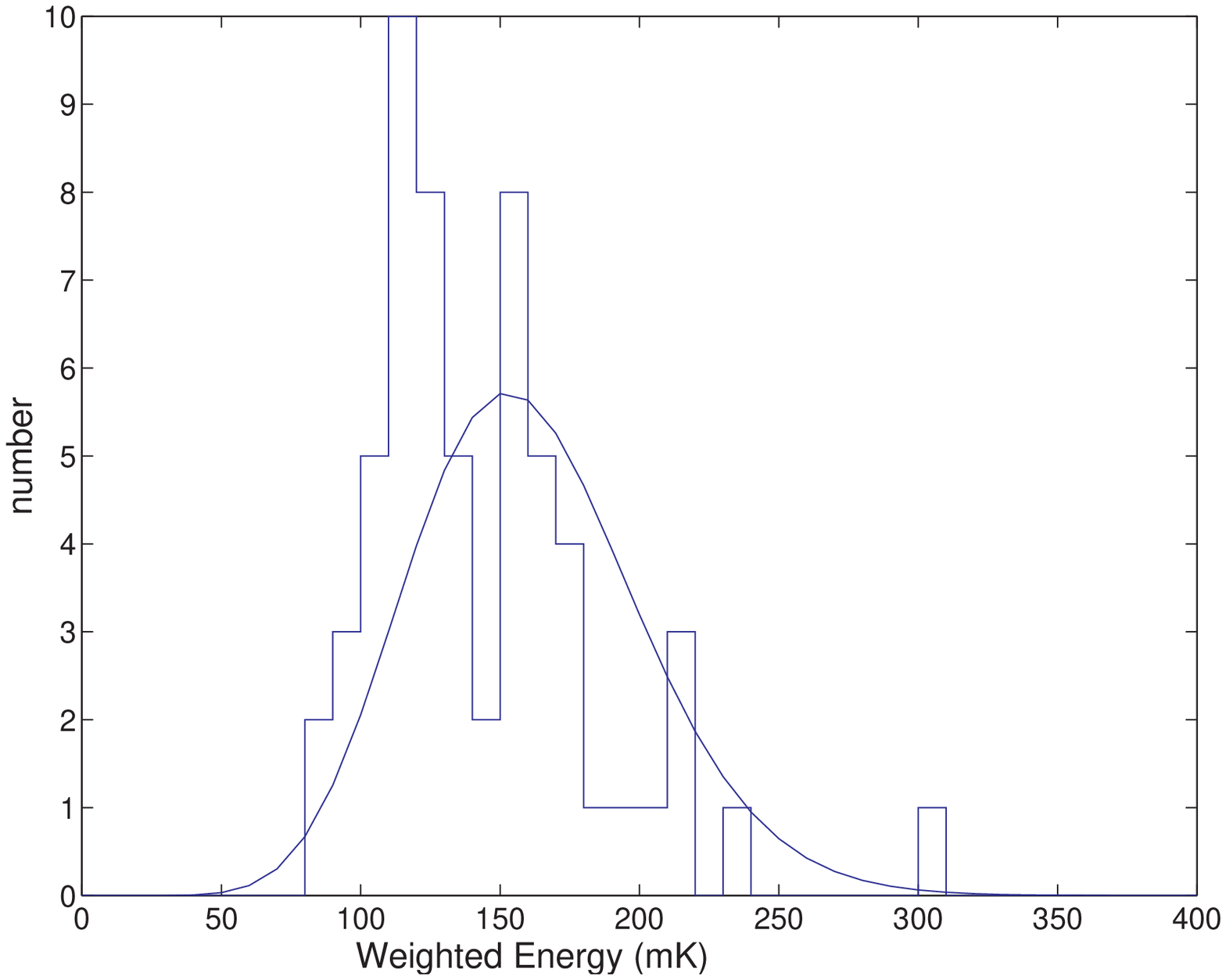 scaled 500}}
   \caption{The histogram of the weighted burst energies for the calibration series
cal2\_327
   and the distribution function of Eq.~(\protect\ref{chi4}).}
   \label{chi4dof}
\end{figure}

Both the distribution for the mode burst energy and the weighted burst energy are
described by only two parameters, the size of the signal and the noise temperature of the
detector.  That the actual data from the detector follows these distributions is shown in
Fig.~\ref{chi} and Fig~\ref{chi4dof}.

\begin{figure}
  	\centerline{\BoxedEPSF{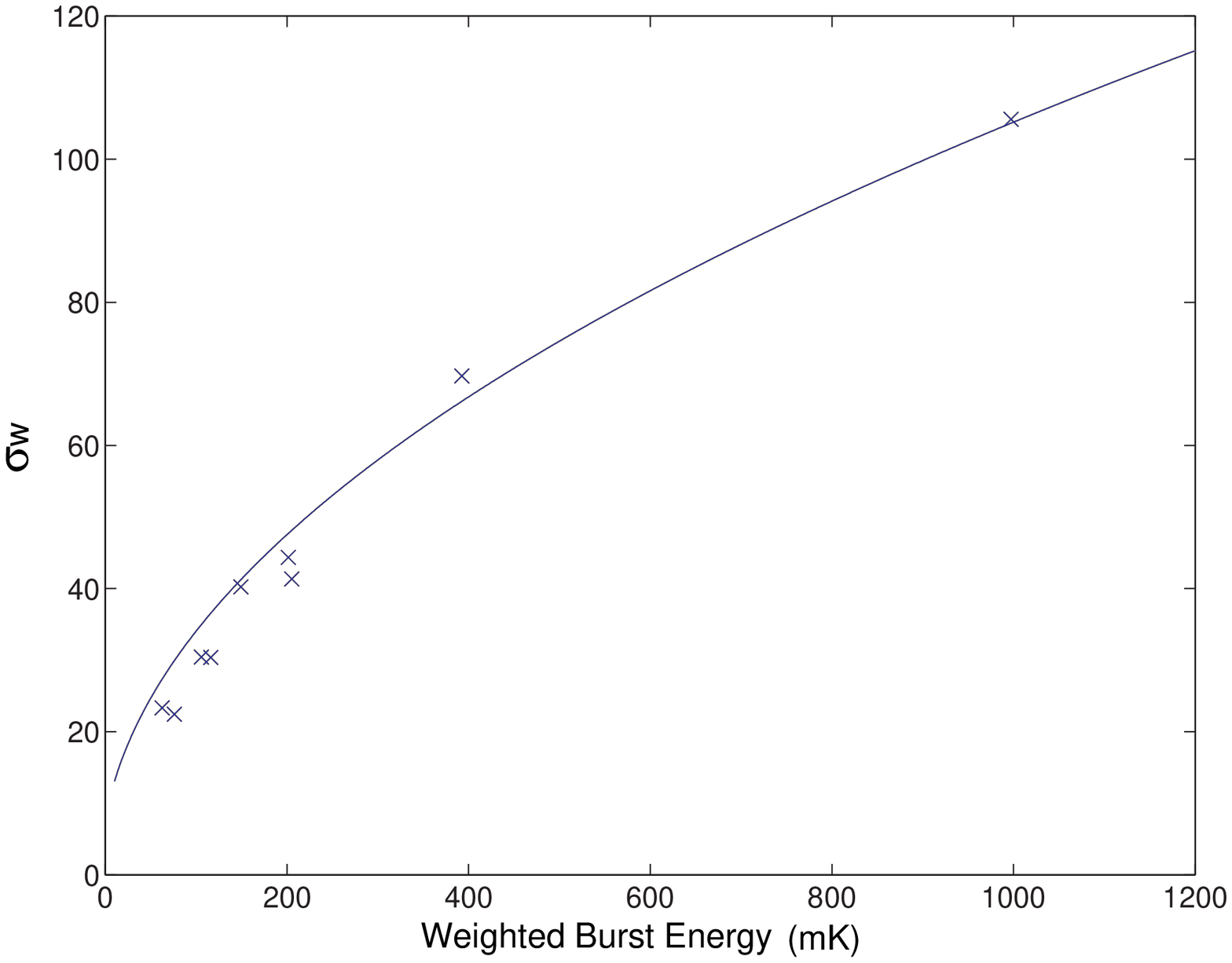 scaled 500}}
   \caption{The spread in the burst energy of a signal due to stationary noise.  Each data
   point is the standard deviation of the energy estimates for a calibration series.  The
solid
   line is a theoretical curve generated from the square root of
   Eq.~(\protect\ref{energyvariance}) with a noise temperature of 5.5 mK.}
   \label{eunc}
\end{figure}

\begin{figure}
  	\centerline{\BoxedEPSF{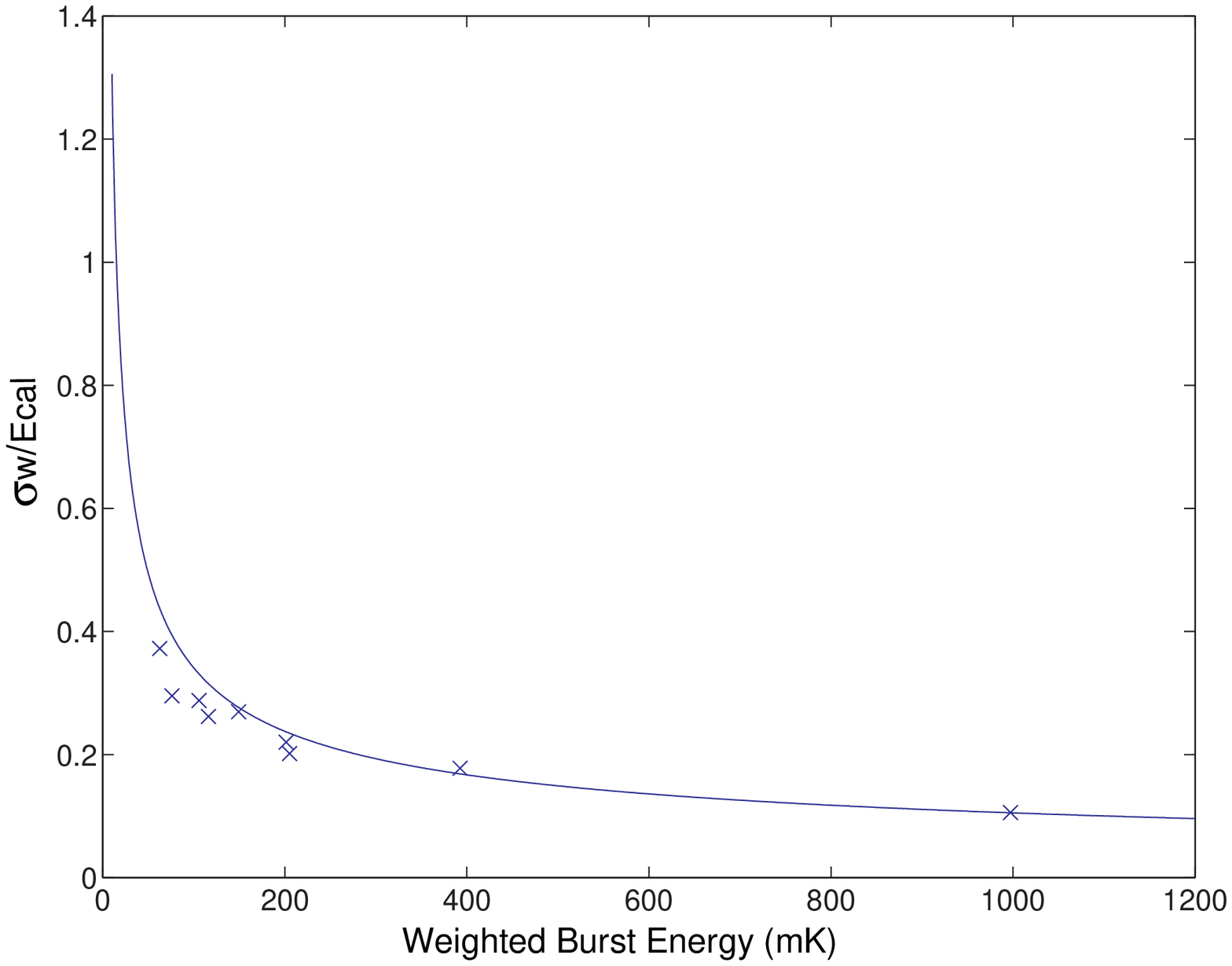 scaled 500}}
   \caption{ The fractional spread in burst energy of a signal mixed with stationary
noise.
   This is simply the results of Fig.~\protect\ref{eunc} divided by the mean value of the
burst
   energy.} 
   \label{func}
\end{figure}

Equation (\ref{energyvariance}) is the important result with regards to a coincidence
search.  If we identify the spread in energy due to the interaction with the noise as the
square root of Eq.(\ref{energyvariance}), call it $\sigma_{w}$, then for a given noise
temperature the spread increases as the square root of the signal strength. 
Figure~\ref{eunc} demonstrates that the data from Allegro matches the theory well.  This
curve is used to define the window of a coincidence in energy.  Although the spread
increases with increasing signal strength, the fractional change in energy, defined as
$\sigma_{w}/E_{cal}$ decreases as
$1/\sqrt{E_{cal}}$ as is shown in Fig.~\ref{func} .  

\section{Non-Stationary Noise}

\begin{figure}[h]
  	\centerline{\BoxedEPSF{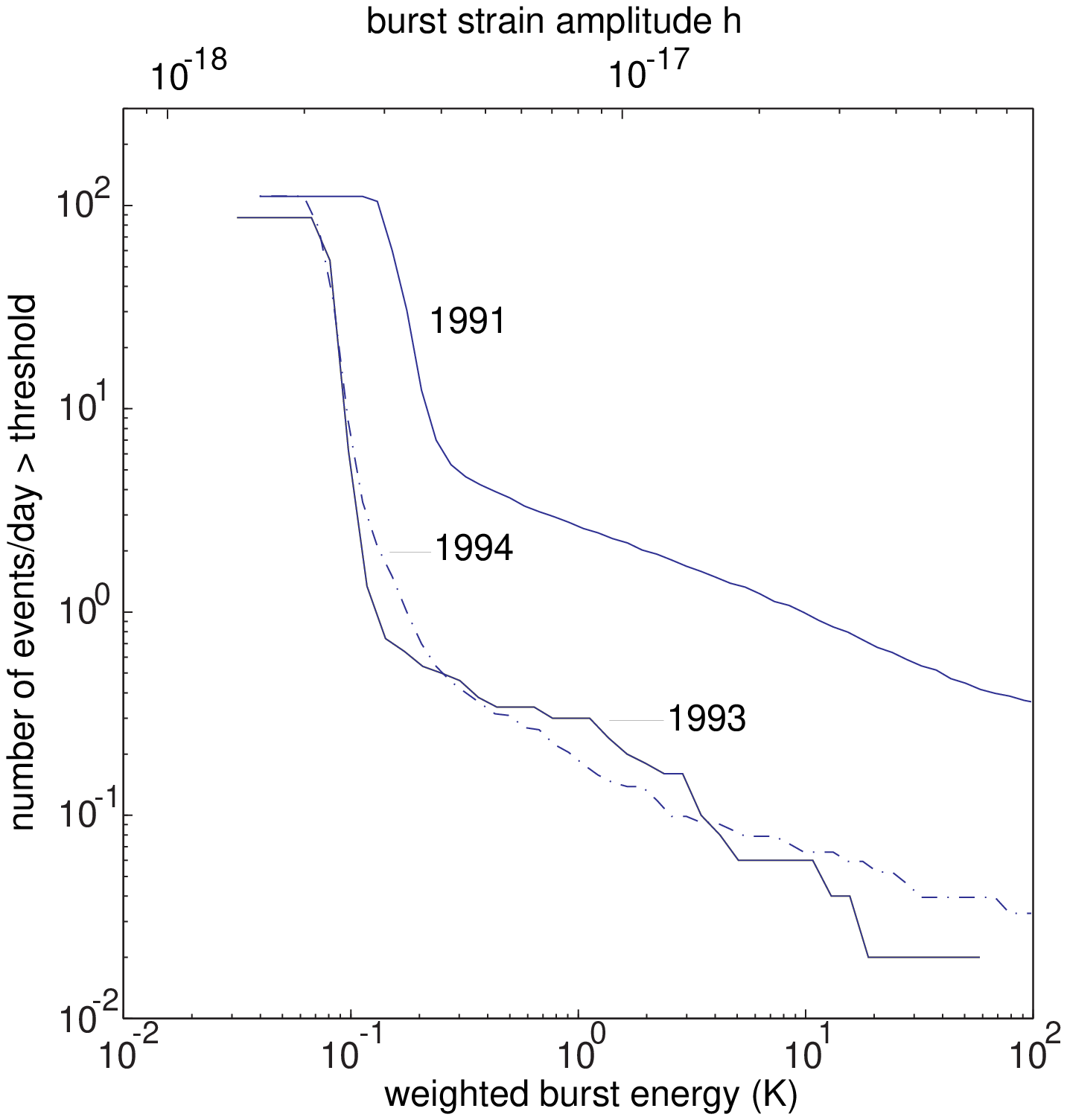 scaled 500}}
   \caption{Allegro energy spectrum for 1991, 1993 and 1994.  The bottom scale gives the
signal
   threshold in Kelvin, the top scale gives the signal threshold in terms of the burst
strain amplitude of a gravity wave incident with optimum polarization and direction.}
   \label{spec}
\end{figure}

The previous section dealt with the effects of Stationary noise on the accuracy of the
event parameters time and energy.  There is another class of noise, non-Stationary noise,
which effects the running of the detector as an observatory.  Figure~\ref{spec} shows the
Allegro energy spectrum from 1991, 1993 and 1994.  Each spectrum is divisible in two
parts, the low energy stationary noise and the background events which could be from any
number of mechanisms related to the detector or surroundings, or could be from gravity
waves.  The lowering of the background from 1991 to 1993 is attributed to two causes. 
First, the antenna was warmed to 15 K at the beginning of 1993 which may have reduced
trapped flux in the superconductors or released some built up mechanical stress.  Second,
there were a number of background sources identified after 1991: millisecond electrical
transients, earthquakes from around the globe, and buses hitting a pothole outside the
Physics Building.  Examining the raw data associated with each event outside of the
thermal distribution allowed events produced by these sources to be easily recognized and
vetoed with only a slight increase in the detector dead time.

\section{Conclusion}

We have described the data acquisition and analysis procedures of the Allegro gravity wave
detector.  The creation of an optimal filter to look for burst signals was discussed in
detail, and the uncertainties in assigning a time and energy to an event due to stationary
noise were calculated and shown to match the data, setting the windows for both
quantities in coincidence searches. 

This research is supported by the National Science Foundation under
Grant  No. \mbox{PHY-9311731}

\bibliographystyle{prsty}

\end{document}